\documentclass{mem}
\usepackage{natbib}\usepackage{balance}
\usepackage{epsfig}
\usepackage[a4paper]{hyperref}
\idline{75}{1}
\begin{document}
\def\teff{$T\rm_{eff }$}
\def\kms{$\mathrm {km s}^{-1}$}

\title{
Evolutionary Paths for Galaxies and AGNs:
New Insights by the Spitzer Space Telescope
}

\subtitle{}

\author{
A. \,Franceschini\inst{1}, G. \,Rodighiero\inst{1}, S. \,Berta\inst{1}
\and P. \, Cassata\inst{1,2}
          }

  \offprints{A. Franceschini}

\institute{
Dipartimento di Astronomia --
Universita' di Padova, Vicolo Osservatorio 2,
I-35122 Padova, Italy,  \email{franceschini@pd.astro.it}
\and
IASF/INAF, Milano
}

\authorrunning{A. Franceschini et al.}

\titlerunning{IR Evolutionary Paths for Galaxies and AGNs
}

\abstract{
We compare the history of the galaxy mass build-up, as inferred from near-IR observations, 
and the Star Formation Rate of massive stars in the comoving volume traced by deep 
extensive far-IR surveys, both possible now with the Spitzer Space Telescope.
These two independent and complementary approaches to the history of
galaxy formation consistently indicate that a wide interval of cosmic epochs
between $z\sim0.7$ to $\sim2$ brackets the main evolutionary phases.
The rate of the integrated galaxy mass growth indicated by the IR-based comoving 
SFR appears consistent with the observed decrease of the stellar mass 
densities with redshift.
There are also indications that the evolution with $z$ of the total 
population depends on galaxy mass, being stronger for moderate-mass, but 
almost absent up to $z=1.4$ for high-mass galaxies, thus confirming 
previous evidence for a "downsizing" effect in galaxy formation.
The most massive galaxies appear already mostly in place by $z\sim 1$.

Although a precise matching of this galaxy
build-up with the growth of nuclear super-massive black-holes is not possible 
with the present data (due to difficulties for an accurate census of the 
obscured AGN phenomenon), some preliminary indications reveal a similar 
mass/luminosity dependence for AGN evolution as for the hosting galaxies.

}
\maketitle{}

\section{Introduction}

The cosmological origin of the galaxy morphological sequence can now be very effectively 
constrained by the wide IR multi-wavelength coverage offered by the Spitzer Space Telescope,
including the near-IR photometric imaging by the IRAC and the far-IR imaging by the MIPS 
instruments. By combining these with the unique imaging capabilities of HST/ACS and the enormous 
photon-collecting power of spectrographs on large ground-based telescopes (VLT, Keck),
we have now a definite chance to directly picture {\sl in-situ} the process of galaxy formation.

The latter has remained a rather controversial issue untill recently.
Published results from high redshift galaxy surveys appear not unfrequently
in disagreement with each other (see Faber et al. 2005 for a recent review). 
This is partly due to the small sampled areas and the 
corresponding substantial field-to-field variance. 
However, a more general problem stems from the apparent conflict between reports
of the detection of massive galaxies at very high redshifts (e.g. Cimatti et al. 2004; 
Glazebrook et al. 2004) and indications for a fast decline in the comoving number 
density at $z>1$ (e.g. Franceschini et al. 1998; Fontana et al. 2004).

The most direct way of constraining the evolutionary history of galaxies and trying
to resolve these discrepancies is to derive the redshift-dependent stellar mass 
functions from deep unbiased surveys (e.g. Bundy, Ellis \& Conselice 2005).
We contribute to this effort by exploiting in this paper very deep public 
imaging by the GOODS project to select a complete sample of high-z 
($z\leq 2$) galaxies selected in the 3.6 $\mu$m IRAC near-IR band. 
Near-IR surveys are best suited for the study of faint high-redshift galaxy 
populations, for various reasons.  Compared to UV-optical selection, 
the observed fluxes are minimally affected by dust extinction. At the same 
time they are good indicators of the stellar mass content of galaxies 
(Dickinson et al. 2003; Berta et al. 2004), and closer to provide a 
mass-selection tool.  
For typical spectra of evolved galaxies, the IRAC channel also 
benefits by a K-correction particularly favourable for the detection of
high-redshift galaxies.

In addition to the Spitzer observations, the GOODS and related CDFS projects 
have provided the community with an unprecedented amount of high quality
optical and near-IR data in CDFS, particularly the very deep multi-band ACS 
imaging, allowing the most accurate morphological analysis currently possible.
We will illustrate here the power of combining 
such multi-wavelength information in the analysis of the evolutionary
mass and luminosity functions of faint galaxies.


A complementary view on galaxy formation is possible by direct sampling the rate of star formation based 
on suitable tracers. As established by exploratory observations with ISO and SCUBA (e.g. Franceschini et al.
2001; Elbaz et al. 2002; Smail et al. 2002), most accurate SFR determinations require the detection of 
dust emission in the far-IR, which often includes the majority of the radiant energy by massive stars.
Our direct estimates of the evolutionary stellar mass functions are then compared with the most updated
results on the comoving density of star formation by Spitzer far-IR observations.
Modulo an assumption about the stellar initial mass function, the two approaches provide
the complementary integral and differential views of star formation, respectively.

Galaxy formation has involved a variety of complex phenomena (see e.g. Baugh et al. 2005), 
including tidal interactions and merging, black-hole formation, accretion, and feedback
processes. The ubiquitous presence
of super-massive black-holes in galaxy nuclei (Magorrian et al. 1998) and the
relevant observed scaling laws (Ferrarese and Merritt, 2000) suggest that 
feedback by nuclear activity (Springel et al. 2005) might have had an important
influence on the process.
This motivates us in looking also for comparison with the evolutionary patterns in the Active 
Galactic Nuclei population.

We adopt a standard cosmology ($\Omega_M$=0.3, $\Omega_{\Lambda}=0.7$)
and express the $H_0$ dependence in terms of $h\equiv H_0/100\ Km/s/Mpc$.

\begin{figure*}
\begin{center}
\psfig{file=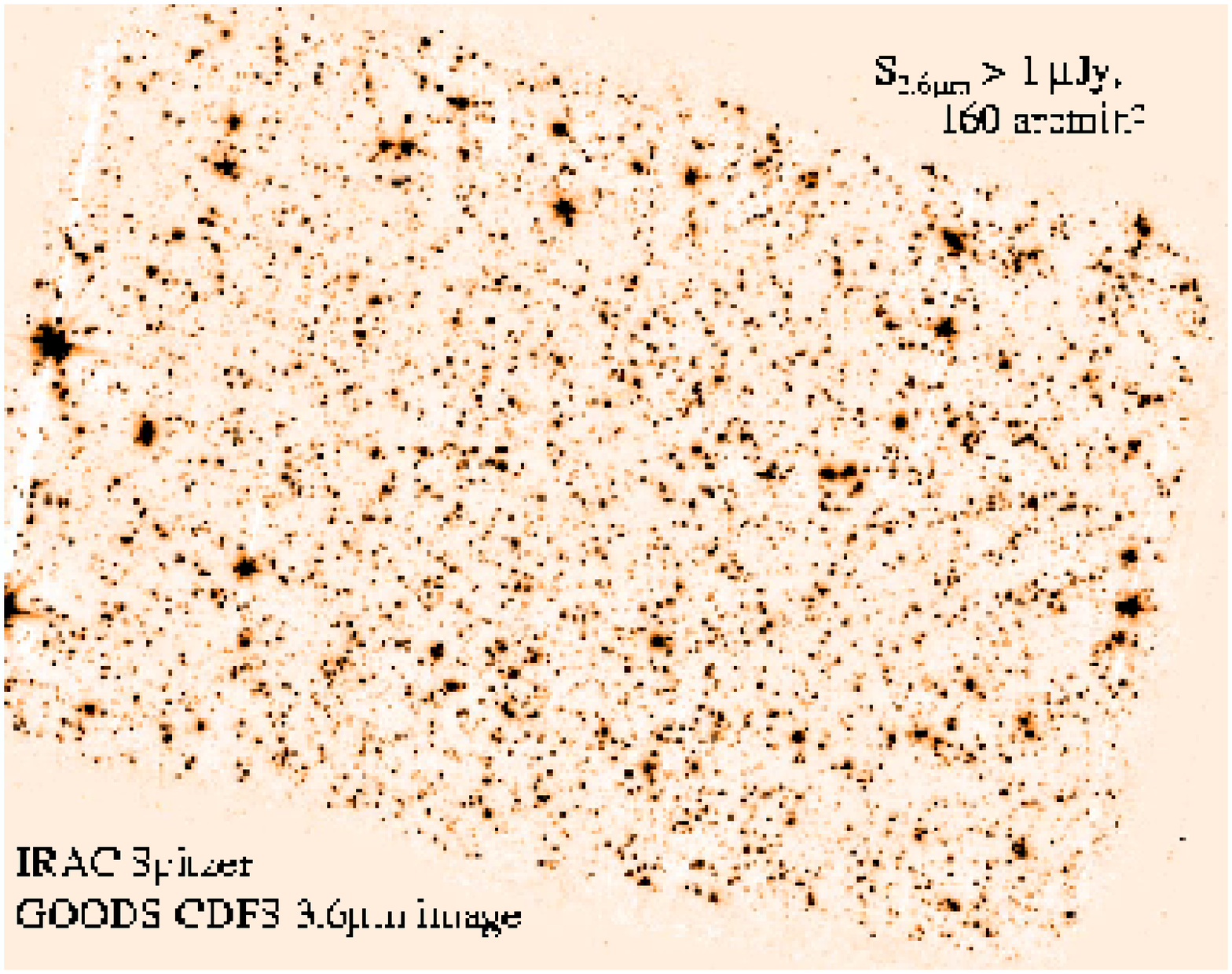,height=7cm,angle=0}
\end{center}
\caption{\footnotesize
Image at 3.6 $\mu$m of the GOODS region in the CDFS. The area covered is 
$12\times 18$ square arcminutes approximately and the exposure time per sky pointing 
was 23 hours as a minimum. The image sensitivity is better than 1 $\mu Jy$ for 
point sources. 5622 sources with $S_{3.6\mu} > 1 \mu Jy$ are detected inside the 
Spitzer/ACS common area of 160 sq. arcmin, 5302 of which are galaxies and 320 stars.
}
\label{f4}
\end{figure*}

\section{Cosmic Evolution of the Galaxy Mass and Luminosity Functions }

Of the two approaches to the history of galaxy formation -- the one based on the study 
of the instantaneous SFR with suitable indicators, and the other 
estimating the evolutionary stellar mass functions -- the latter has been 
recently recognized to benefit by various advantages (e.g. much lower 
extinction uncertainties) and to be more robust.

The evolution of the galaxy stellar mass function has been recently 
studied by us using a multi-wavelength dataset in the Chandra Deep Field South
(CDFS) area, obtained from GOODS (Dickinson et al. 2004) and other projects (VVDS, 
Le Fevre et al. 2004), and including very deep high-resolution imaging by HST/ACS. 
Our reference catalogue of faint high-redshift galaxies, which we 
have thoroughly tested for completeness and reliability, comes from
a deep ($S_{3.6}\geq 1\ \mu$Jy) image by IRAC on the Spitzer Observatory 
(see Figure \ref{f4}).
These data are complemented with extensive optical
spectroscopy by the ESO FORS2 and VIMOS spectrographs, while
deep K-band VLT/ISAAC imaging is also used to derive further complementary 
statistical constraints and to assist the source identification 
and SED analysis.    We have selected a highly reliable
IRAC 3.6$\mu$m sub-sample of 1478 galaxies with $S_{3.6}\geq 10\ \mu$Jy, 
47\% of which have spectroscopic redshift, while for the remaining objects 
both COMBO-17 and $Hyperz$ are used to estimate the photometric redshift.

\begin{figure*}
\begin{center}
\psfig{file=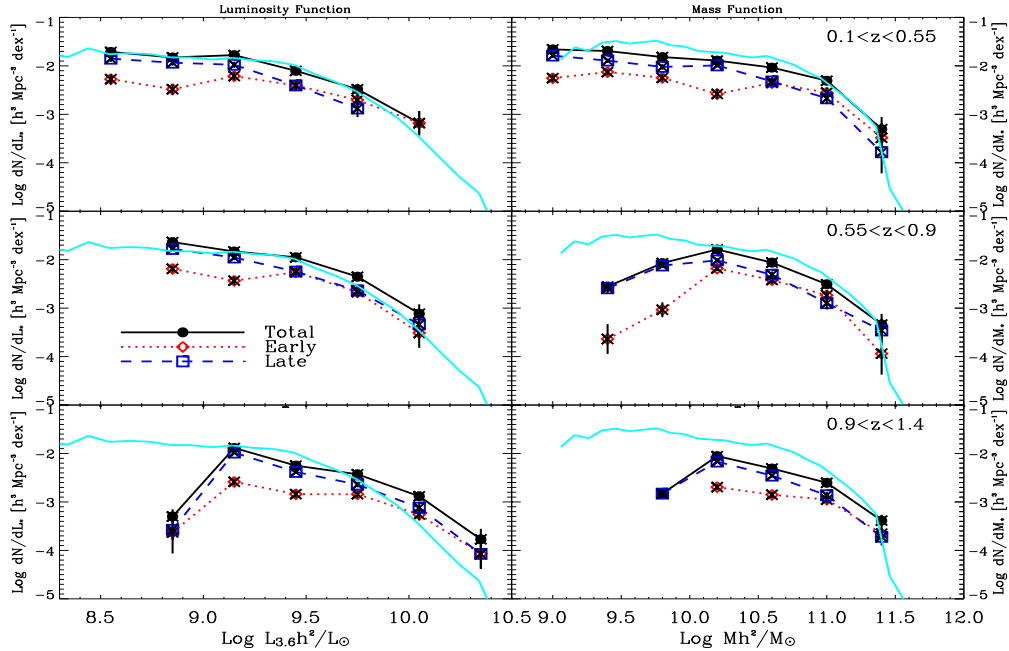,height=9cm,width=13.5cm,angle=0}
\end{center}
\caption{\footnotesize
Mass ({\sl right-hand panels}) and luminosity ({\sl left-hand panels}) 
function estimates derived from the 3.6 $\mu$m IRAC/GOODS
sample with $S_{3.6}>$10 $\mu$Jy, splitted into three redshift
bins from $z=0.1$ up to $z=1.4$. The contributions of the various
morphological classes is marked with different symbols:
early-types (open diamonds - dotted lines), late-types (open squares
- dashed lines), total (filled circles - solid lines). The thin solid 
line on the right marks the local mass function from Cole et al. (2001). 
}
\label{lf_mf}
\end{figure*}

Based on this extensive dataset, we have estimated time-dependent
galaxy luminosity and stellar mass functions, while luminosity/density evolution 
is further constrained with the number counts and redshift distributions.
The deep ACS imaging allows us to differentiate these evolutionary paths by 
morphological type, which can be reliably performed at least up 
to $z\sim 1.5$ for the two main early- (E/S0) and late-type (Sp/Irr) classes.

These data, as well as our direct estimate of the stellar
mass function above $M_\ast h^2=10^{10} M_\odot$ for the spheroidal subclass, 
consistently evidence a progressive dearth of such objects to occur
starting at $z\sim 0.7$, paralleled by an increase in luminosity.
A similar trend, with a more modest decrease of the mass function,
is also shared by spiral galaxies, while the irregulars/mergers
show a positive evolution (increased number) up to $z\simeq 1.5$.

It is interesting to note that the decrease of the comoving density with redshift of the
total population appears
to depend on galaxy mass, being stronger for moderate-mass, but almost
absent until $z=1.4$ for high-mass galaxies, thus confirming 
previous evidence for a "downsizing" effect in galaxy formation (e.g. Cowie et al. 1996).

\begin{figure*}[t!]
\begin{center}
\psfig{file=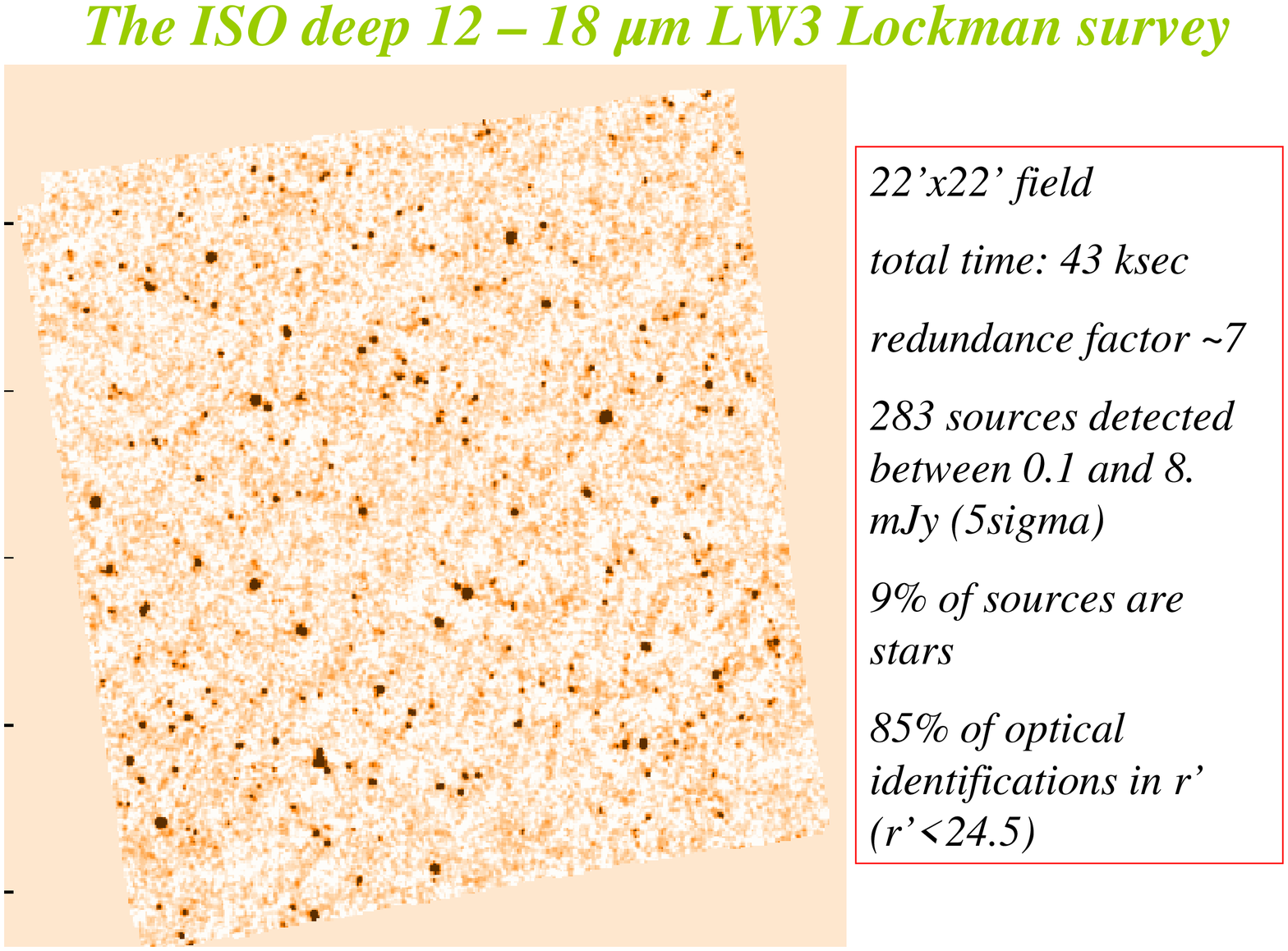,height=7cm,angle=0}
\end{center}
\caption{\footnotesize
Image at 15 $\mu$m of an area of 22x22 arcmins in the Lockman Hole by Rodighiero et al. (2004).
Details about the observations are reported in the figure.
}
\label{f1}
\end{figure*}

\section{Probing the Evolution of the Comoving Star Formation Rate with Far-IR Observations}

The usual way of determining the star formation activity in high-redshift galaxies is based on 
optical observations of the UV rest-frame flux (Lilly et al. 1995; Madau et al. 1996; Le Fevre 
et al. 2004).  However, it has become recently evident that a probably major fraction of 
the emission by the most massive, luminous and short-lived stars, during the stage 
when they are still embedded inside their
parent dusty molecular clouds, is optically extinguished and reprocessed at wavelengths from 
the mid-IR to the sub-millimeter.

\subsection{The ISO and SCUBA explorations}

The Infrared Space Observatory mission and the SCUBA/JCMT observations have
offered the first tools for systematic cosmological surveys at long wavelengths.
The good imaging capabilites of ISO have been exploited for deep mid-IR 
observations with sensitivities sufficient to detect sources at cosmological redshifts
(Elbaz et al. 1999, 2002; Franceschini et al. 2001).
Surveys in the wide band at 12-18$\mu$m ($\lambda_{eff}=15\ \mu$m) have been performed 
in the ISOCAM GT, over a total area of 1.5 square degrees, with $>$1000 sources detected
(Elbaz et al. 1999). 
Rodighiero et al. (2004) and Fadda et al. (2004) report a deep and a shallow 
survey of the Lockman Hole. An image of the deep field is reported in Figure \ref{f1}.
The two Hubble Deep Field areas (North and South, total of $\sim 50$ sq. arcmin)
have been deeply surveyed at 15 $\mu m$ down to 100 $\mu Jy$ (Rowan-Robinson et al. 1997; 
Aussel et al. 1998). 
At brighter fluxes, ELAIS has observed a total of 12 square degrees at 15 $\mu m$ 
(Oliver et al. 2000; Vaccari et al. 2005).

\begin{figure}
\psfig{file=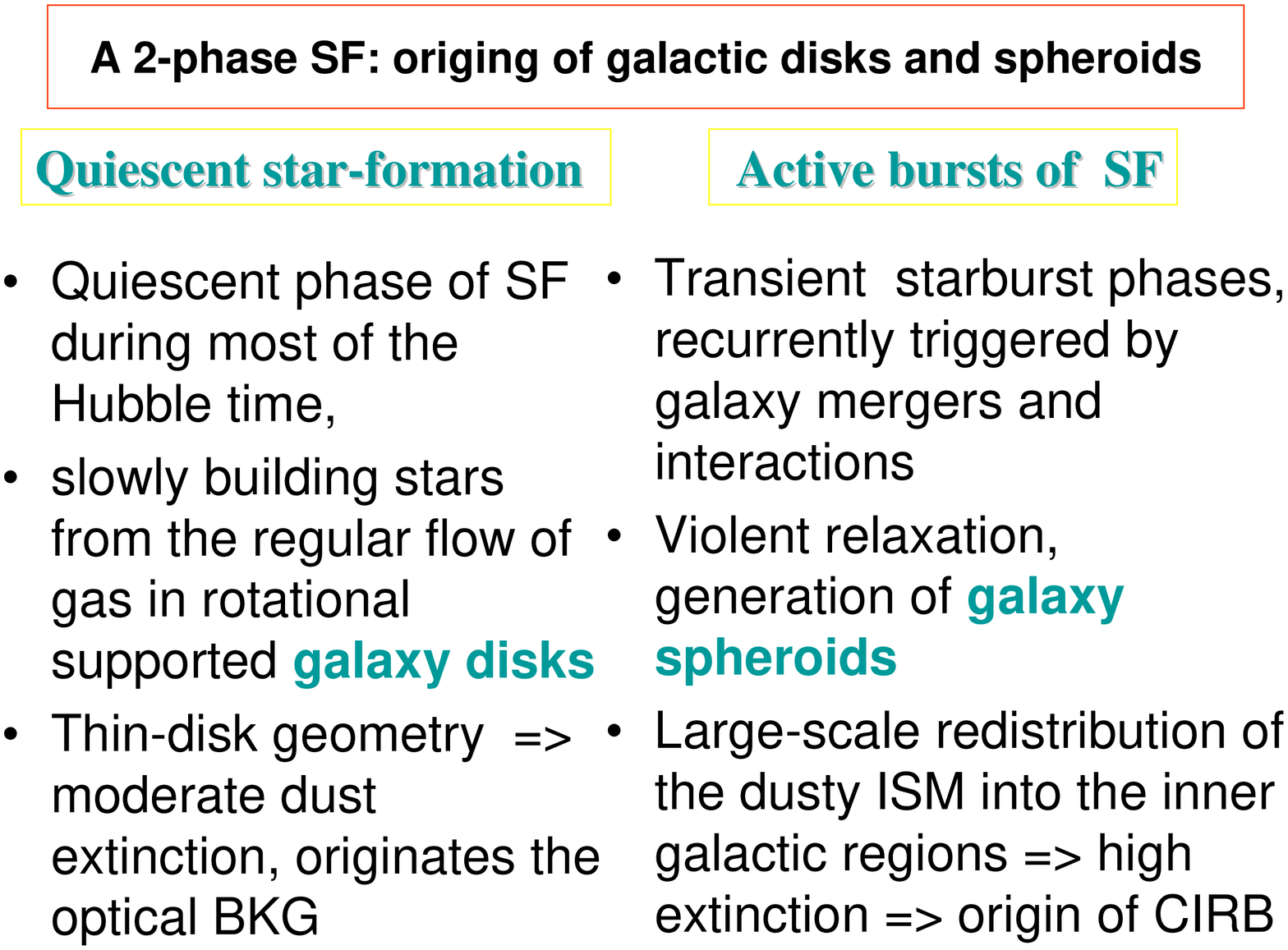,height=5cm,width=6.5cm,angle=0}
\caption{\footnotesize
A two-phase scheme for the evolution of galaxies selected in the optical and the IR.
}
\label{f2}
\end{figure}

\begin{figure*}
\begin{center}
\psfig{file=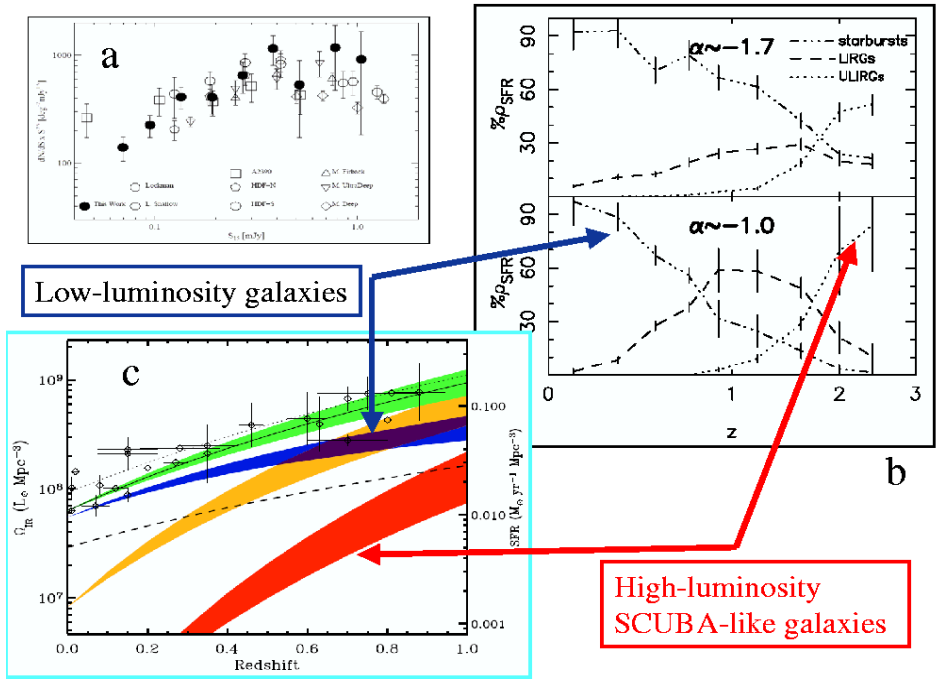,height=9cm,width=12.5cm,angle=0}
\end{center}
\caption{\footnotesize
 {\sl Panel a}: comparison of the differential 15 $\mu$m galaxy counts measured by ISO
and Spitzer (Teplitz et al. 2005), showing excellent agreement.  {\sl Panel b}: 
relative contribution of low luminosity starbursts ($L_{IR} <10^{11}\ L_\odot$), IR luminous
sources (LIRGs, $L_{IR}>10^{11}$) and ULIRGs ($L_{IR} <10^{11}$) to the total SFR
density of the universe as a function of redshift (Perez-Gonzales et al. 2005).
{\sl Panel c}: evolution of the comoving IR galaxy emissivity up to z=1 (green region) and the 
respective contributions from low luminosity galaxies ($L_{IR} <10^{11}$, blue), IR luminous 
sources ($L_{IR}>10^{11}$, orange) and ULIRGs ($L_{IR} <10^{11}$, red).
}
\label{f3}
\end{figure*}

The differential 15$\mu$m counts of extragalactic sources 
revealed a sudden upturn at $S_{15}< 3\ mJy$, and a later convergence 
below $S_{15}\sim 0.3\ mJy$ (Elbaz et al. 1999; Gruppioni et al. 
2003). These counts, far in excess of the no-evolution predictions, 
required strong evolution of the galaxy luminosity functions.
%
%
At the faintest mid-IR fluxes, the redshift distributions of the ISO galaxy population 
showed an excess number of luminous sources between z=0.5 and z=1. The upper redshift
boundary is due to the ISOCAM mid-IR camera responsivity, implying a strong
K-correction penalty for $z>1$.

These results have been analyzed by Franceschini et al. (2001) and Elbaz et al. (2002) 
in conjunction with data and limits on the integrated sky intensity between 1 and 1000 
$\mu$m set by COBE observations (the CIRB background), and with 
the sub-mm galaxy counts at 850 $\mu$m by SCUBA (see review by Smail et al. 2002).
The combined multi-wavelength constraints imply that the far-IR volume emissivity 
of galaxies increases strongly from local to redshift $z\sim 1$ and should flatten
or converge above.  A comparison of the results of optical and IR cosmological surveys 
has been interpreted by Franceschini et al. (2001) to indicate that galaxies during
their life form stars both in a quiescent mode, mostly responsible for the optical emission,
and through short-lived starbursting episodes originating the far-IR emission, as
summarized in Figure \ref{f2}.

\begin{figure}
\begin{center}
\psfig{file=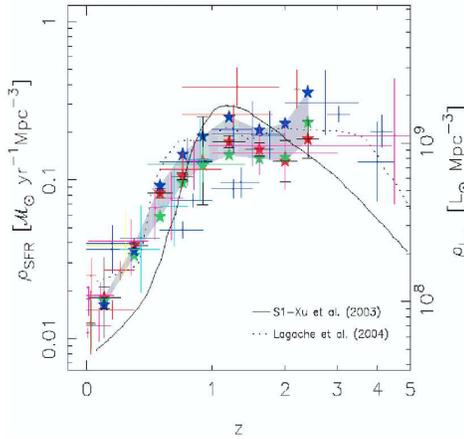,height=6cm,width=7.5cm,angle=0}
\end{center}
\caption{\footnotesize
Evolution of the comoving IR luminosity density and SFR as a function of redshift, 
taken from Perez-Gonzales et al. (2005). All values in the figure are computed for $h=0.7$.
}
\label{eta}
\end{figure}

\begin{figure}
\psfig{file=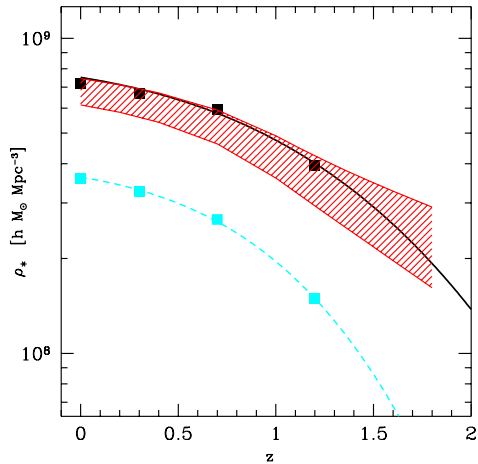,height=6.5cm,width=6.5cm,angle=0}
\caption{\footnotesize
The red-shaded area shows the expected evolution of the total integrated stellar mass 
density, as inferred from measurements of the comoving SFR. The upper datapoints and the 
continuous line are direct estimates by Franceschini et al. (2006) for the total 
galaxy population. The lower (cyan) points are the corresponding measurement for 
spheroidal galaxies, showing a somewhat faster convergence with $z$.
}
\label{Mint}
\end{figure}

\subsection{Results by the Spitzer Space Telescope }

The far-IR imagers (MIPS and IRS) on Spitzer not only have confirmed the ISO 
results (Teplitz et al. 2005, see panel {\sl (a)} in Figure \ref{f3}), 
but, thanks to the longer wavelength capabilities of the
24 $\mu$m channel, have been able to extend the ISO surveys to higher redshifts
cover the critical redshift interval of $z\sim1$ to $\sim2$.

A region of the CDFS was observed at 24 $\mu$m by Le Floc'h et al. (2005) during the 
MIPS instrument GTO over a total area of 0.6 $deg^2$ and PSF FWHM of 6 arcsec.
A sample of 2600 sources brighter than 80 $\mu Jy$ was combined with existing optical
data in the field and used to derive bolometric IR luminosity functions and SFR's
from z=0 to $\sim1$. These results imply that the comoving IR energy density of the 
Universe evolves like $(1+z)^{3.9 \pm 0.4}$ up to $z\simeq 1$ (panel {\sl (c)} in Fig. 
\ref{f3}).

From MIPS 24 $\mu$m observations of the CDFS and HDFN, complemented with a systematic 
photometric redshift analysis exploiting the Spitzer IRAC data, Perez-Gonzales et al.
(2005) confirm on one side the fast increase of the comoving SFR density to $z=0.8$. 
As shown in Figure \ref{eta},
the SFR density is found to continue rising from $z\sim0.8$ to 1.2 with a smaller slope,
but then to remain roughly constant above, in quite good agreement with results by 
Franceschini et al. (2001).

\section{Discussion}

It is of interest to verify how consistent the results about the redshift evolution of 
the integrated comoving stellar mass density $\rho_M(z)$ in Fig. \ref{lf_mf} and the
integrated star formation rate 
$\rho_{SFR}(z)$ in Fig. \ref{eta} are. Both quantities are based on the assumption of 
a universal Salpeter-like stellar IMF. The relationship of the two is simply:
\begin{equation}
\rho_M(z) = \rho_M(0) - \int_0^z \rho_{SFR}(z) {dt \over dz} dz
\end{equation}
where $\rho_{SFR}(z)\simeq 0.014 \ h \ (1+z)^{3.9}$ at $z<1$ and constant above (from 
Perez-Gonzales et al.), and where we can approximate $dt/dz=1/[H_0 (1+z)^{2.5}]$. Assuming 
$\rho_M(0)\simeq (6.8\pm 0.8)10^8\ h \ M_\odot/Mpc^3$ from Fukugita et al. (1998),
we derive the evolution of the integrated stellar density reported as red shaded region 
in Figure \ref{Mint}. This is compared with the direct estimates by Franceschini et al.
(2006) for the total galaxy population (suitably scaled from their fig.19 to correct
for the different mass boundary). Within the large uncertainties, we see here
that the two complementary views of the history of star formation appear to provide
consistent results, hence in broad agreement with the assumed Salpeter IMF.

Panels {\sl (b)} and {\sl (c)} in Fig. \ref{f3} also reveal some interesting features
in the z-evolution of the SFR density as a function of the galaxy luminosity.
Low-luminosity sources display their main SF activity below $z=1$, 
while the highest-L ULIRGs develop their main activity phases at $z>1.5$. 
Intermediate luminosity objects peak at intermediate redshifts of 0.7 to 1.5. 
This, together with the results illustrated in Fig. \ref{lf_mf}, provide concordant evidence 
that the formation and evolution of galaxies has been a strong function of luminosity and mass,
as anticipated and discussed by various authors (Cowie et al. 1996, Franceschini et al.
1998, 1999; Treu et al. 2005).

Various attempts have been proposed to explain this effect, which seems to indicate a sort
of inversion 
in the process of dynamical assembly of galaxies compared to the hierarchical expectation. 
One possibility considers energy feedback by a nuclear AGN stopping
gas accretion and star formation in the most massive systems at high-$z$ (Granato
et al. 2004; Springel et al. 2005), while leaving it undisturbed in the lower mass galaxies
where AGN activity is irrelevant.

\begin{figure}
\begin{center}
\psfig{file=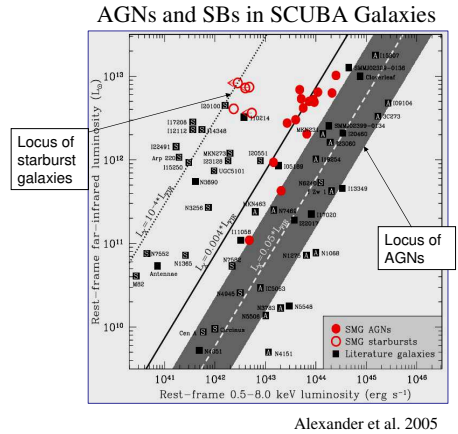,height=6cm,width=8.cm,angle=0}
\end{center}
\caption{\footnotesize
Relation of far-IR to X-ray luminosity in high-z SCUBA galaxies (from Alexander et al. 2005).
The bulk of these (red circles) are found to require both starburst and AGN 
simultaneous emissions.
}
\label{X}
\end{figure}

This interpretation would call for a close relationship of the evolutionary 
histories of star-formation and nuclear Black Hole Accretion in AGNs.
Unfortunately, a precise matching of galaxy build-up and the growth of 
nuclear super-massive black-holes is not possible 
with the present data, due to difficulties for an accurate census of the 
obscured AGN phenomenon.  The problem has not yet been finally settled 
by the inclusion of data by the Spitzer surveys, bacause even a detailed IR
multi-wavelength coverage is not sufficient to identify the whole of the
obscured AGN population (see Franceschini et al. 2005; Polletta et al. 2006). 

Preliminary results, mostly from X-ray surveys, confirm on one side that 
AGN and starburst activity occur concomitant in luminous high-z forming galaxies 
(as shown e.g. in the remarkable outcome by Alexander et al. 2005 
reported in Figure \ref{X}). 

It is also interesting to note that a similar mass/luminosity dependence for AGN 
evolution as for the hosting galaxies 
(with fast evolution with cosmic time for the highest-L and deferred
activity for the lower-L objects)
has been found in the luminosity functions of type-1 AGNs
derived from complete X-ray surveys by Hasinger et al. (2005).



\bibliographystyle{aa}

\end{document}